\documentclass[12pt]{article}
\usepackage{amssymb}
\usepackage{amsmath}

\begin{document}

\title{Separability and Entanglement-Breaking \\
in Infinite Dimensions}
\author{A. S. Holevo\thanks{
Steklov Mathematical Institute, Moscow, Russia.} \thanks{
Part of this work was done when the author was visiting the Institute of
Mathematical Physics, TU Braunschweig in frame of the A. von Humboldt
Foundation Follow-up Program.}, M. E. Shirokov$^{*}$ and R. F. Werner\thanks{
Institute of Mathematical Physics, TU Braunschweig.}}
\date{}
\maketitle

\section{Introduction}

In this paper we give a general integral representation for
separable states in the tensor product of infinite dimensional
Hilbert spaces and provide the first example of separable states
that are not countably decomposable. We also prove the structure
theorem for the quantum communication channels that are
entanglement-breaking, generalizing the finite-dimensional result of
M. Horodecki, Ruskai and Shor. In the finite dimensional case such
channels can be characterized as having the Kraus representation
with  operators of rank 1. The above example implies existence of
infinite-dimensional entanglement-breaking channels having no such
representation.

\section{Separable states}

In what follows $\mathcal{H},\mathcal{K},\dots $ denote separable
Hilbert spaces; $\mathfrak{T}(\mathcal{H})$ denotes the Banach space
of trace-class operators in $\mathcal{H}$, and
$\mathfrak{S}(\mathcal{H})$ -- the convex subset of all density
operators. We shall also call them \textit{states }for brevity,
having in mind that a density operator $\rho $ uniqiely determines a
normal state on the algebra of all bounded operators in
$\mathcal{H}$. Equipped with the trace-norm distance,
$\mathfrak{S}(\mathcal{H})$ is a complete separable metric space. It
is known\textbf{\ } \cite{Dev}, \cite{D-A} that a sequence of
quantum states converging to a state in the weak operator topology
converges to it in the trace norm.

If $\pi $ is a Borel probability measure on $\mathfrak{S}(\mathcal{H}),$ the
relation
\begin{equation}
\bar{\rho}(\pi )=\int_{\mathfrak{S}(\mathcal{H})}\sigma \pi (d\sigma ),
\label{bary}
\end{equation}
where the integral exists as Bochner integral, defines a state
called \textit{barycenter} of $\pi$. Let $\mathcal{P}$ be a set of
Borel probability measures on \textit{$\mathfrak{S}(\mathcal{H}).$}
Recall that weak convergence of probability measures means
convergence of integrals of all continuous bounded functions (see e.
g. \cite{Par}). By using the above mentioned result in \cite{Dev},
\cite{D-A} it is easy to see that the map $\pi \mapsto
\bar{\rho}(\pi)$ from $\mathcal{P}$ onto $\mathfrak{S}(
\mathcal{H})$ is continuous. The following result was established in
\cite {H-Sh-2}:

\textbf{Theorem 1.} \textit{The set $\mathcal{P}$} \textit{\ is weakly
compact if and only if its image $\mathcal{A}$ under the map }$\pi \mapsto
\bar{\rho}(\pi )$ \textit{given by (\ref{bary})} \textit{is a compact subset
of $\mathfrak{S}(\mathcal{H})$.}

The following lemma is an amplified version of the Choquet
decomposition \cite{Alf} adapted to the case of closed convex
subsets of $\mathfrak{S}( \mathcal{H})$. We denote by
$\mathrm{co}\mathcal{A}$ $(\overline{\mathrm{co}} \mathcal{A})$ the
convex hull (closure) of a set $\mathcal{A}$ \cite{J&T}.

\textbf{Lemma 1.} \textit{Let $\mathcal{A}$ be a closed subset of
$\mathfrak{ \ S}(\mathcal{H})$. Then
$\overline{\mathrm{co}}\mathcal{A}$ coincides with the set of
barycenters of all Borel probability measures supported by $
\mathcal{A}$.}

\textbf{Proof.} Let $\rho _{0}\in
\overline{\mathrm{co}}\mathcal{A}$. Then there is a sequence $\{\rho
_{n}\}\subseteq \mathrm{co}\mathcal{A}$ converging to $\rho _{0}$,
so that $\{\rho _{n}\}$ is relatively compact in $
\mathfrak{S}(\mathcal{H})$. The density operator $\rho _{n}$ is
barycenter of Borel probability measure $\pi _{n}$ finitely
supported on $\mathcal{A}$. By the compactness criterion of theorem
1, the sequence $\{\pi _{n}\}$ is weakly relatively compact and thus
has a partial limit $\pi _{0},$ which is supported by the set
$\mathcal{A}$ due to theorem 6.1 in \cite{Par}. Continuity of the
map $\pi \mapsto \bar{\rho}(\pi )=\int \sigma \pi (d\sigma )$
implies that the state $\rho _{0}$ is the barycenter of the measure
$\pi _{0}$.

Conversely, let $\pi$ be an arbitrary probability measure supported
by $ \mathcal{A}$. By theorem 6.3\footnote{ More precisely, it
follows from the construction used in the proof of this theorem.} in
\cite{Par} this measure can be weakly approximated by a sequence of
measures $\pi_{n}$ finitely supported by $\mathcal{A}$. Since $
\bar{\rho}(\pi_{n})$ is in $\mathrm{co}\mathcal{A}$ for all $n$ we
conclude that $\bar{\rho}(\pi)$ is in $\overline{\mathrm{co}}
\mathcal{A}$ due to continuity of the map $\pi \mapsto
\bar{\rho}(\pi)$. $\square$

\textbf{Definition 1.} \textit{A state in}
$\mathfrak{S}(\mathcal{H}\otimes \mathcal{K})$\textit{\ is called
separable if it is in the convex closure of the set of all product
states in} $\mathfrak{S}(\mathcal{H}\otimes \mathcal{K })$.

Lemma 1 implies that separable states are precisely those states
which admit the representation
\begin{equation}
\rho
=\int_{\mathfrak{S}(\mathcal{H})}\int_{\mathfrak{S}(\mathcal{K})}\left(
\rho _{\mathcal{H}}\otimes \rho _{\mathcal{K}}\right) \mu (d\rho
_{\mathcal{H }}d\rho _{\mathcal{K}}),  \label{sepa}
\end{equation}
where $\mu $ is a Borel probability measure on
$\mathfrak{S}(\mathcal{H} )\times \mathfrak{S}(\mathcal{K}).$ In the
finite dimensional case application of Caratheodory's theorem
reduces this to the familiar definition of separable state as finite
convex combination of product states \cite{werner}. If for a
separable state $\rho $ it is possible to find a representation
(\ref{sepa}) with purely atomic $\mu ,$ we call the state
\textit{countably decomposable}. A necessary condition for this is
existence of nonzero vectors $|\alpha \rangle \in \mathcal{H},|\beta
\rangle \in \mathcal{K}$ such that
\begin{equation}
\rho \geq |\alpha \rangle \langle \alpha |\otimes |\beta \rangle \langle
\beta |,  \label{nec}
\end{equation}
cf. \cite{ww}. In Sec. 2 we shall show that there are many separable states
which do not satisfy this condition and hence are not countably decomposable.

In the definition 1 one can replace the set of all product states by
the set of all products of pure states. It is known that the subset
$ \mathfrak{P}(\mathcal{H})$ of pure states (extreme points of
$\mathfrak{S}( \mathcal{H})$) is closed in the trace-norm topology.
The lemma 1 then implies that a state $\rho $ is separable if and
only if there is a Borel measure $ \nu $ on
$\mathfrak{P}(\mathcal{H})\times \mathfrak{P}(\mathcal{K})$ such
that, with some abuse of notation,
\begin{equation}
\rho =\int_{\mathfrak{P}(\mathcal{H})}\int_{\mathfrak{P}(\mathcal{K}
)}|\varphi \rangle \langle \varphi |\otimes |\psi \rangle \langle
\psi |\nu (d\varphi d\psi ).  \label{sep}
\end{equation}

\section{Entanglement-breaking channels}

A \textit{channel} is a linear map $\Phi $:$\mathfrak{T}(\mathcal{H})\mapsto
\mathfrak{T}(\mathcal{H}^{\prime })$ with the properties:

1) $\Phi (\mathfrak{S}(\mathcal{H}))\subseteq
\mathfrak{S}(\mathcal{H} ^{\prime });$ this implies that $\Phi $ is
bounded map and hence is uniquely determined by the infinite matrix
$\left[ \Phi \left( |i\rangle \langle j|\right) \right] ,$ where
$\left\{ |i\rangle \right\} $ is an orthonormal basis in
$\mathcal{H}.$

2) The matrix $\left[ \Phi \left( |i\rangle \langle j|\right) \right] $ is
positive definite in the sense that for a collection of vectors $\left\{
|\psi _{i}\rangle \right\} \subseteq \mathcal{H}$ with finite number of
nonzero elements
\begin{equation}
\sum_{ij}\langle \psi _{i}|\Phi \left( |i\rangle \langle j|\right) |\psi
_{j}\rangle \geq 0.  \label{cp}
\end{equation}

\textbf{Definition 2.} \textit{A channel $\Phi $ is called
entanglement-breaking if for arbitrary Hilbert space }$\mathcal{K}$\textit{\
and arbitrary state} $\omega \in \mathfrak{S}(\mathcal{H}\otimes\mathcal{K})$
\textit{the state} $(\Phi \otimes \mathrm{Id}_{\mathcal{K}})(\omega )$,
\textit{where} $\mathrm{Id}_{\mathcal{K}}$ \textit{is the identity channel
in } $\mathfrak{S}(\mathcal{K})$\textit{, is separable.}

Note that in this definition one can restrict to finite dimensional Hilbert
spaces $\mathcal{K}$. Indeed, an arbitrary state $\omega $ in $\mathfrak{S}(
\mathcal{H}\otimes \mathcal{K})$ with an infinite dimensional $\mathcal{K}$
can be approximated by the states
\begin{equation*}
\omega _{n}=(\mathrm{Tr}(I_{\mathcal{H}}\otimes Q_{n})\omega )^{-1}(I_{
\mathcal{H}}\otimes Q_{n})\omega (I_{\mathcal{H}}\otimes Q_{n}),
\end{equation*}
where $\{Q_{n}\}$ is the sequence of the spectral projectors of the partial
state $\mathrm{Tr}_{\mathcal{H}}\omega $ corresponding to its $n$ largest
eigenvalues. Each state $\omega _{n}$ can be considered as a state in $
\mathfrak{S}(\mathcal{H}\otimes \mathcal{K}_{n})$, where $\mathcal{K}
_{n}=Q_{n}(\mathcal{K})$ is $n$-dimensional Hilbert space. If $(\Phi \otimes
\mathrm{Id)}(\omega _{n})$ is separable for all $n$, then $(\Phi \otimes
\mathrm{Id)}(\omega )$ is also separable as a limit of sequence of separable
states.

The following theorem is a generalization of the result in \cite{R} to the
infinite dimensional case.

\textbf{Theorem 2.} \textit{Channel $\Phi $ is entanglement-breaking
if and only if there is a complete separable metric space
$\mathcal{X}$, a Borel $ \mathfrak{S}(\mathcal{H}^{\prime })$-valued
function $x\mapsto \rho ^{\prime }(x)$ and a positive
operator-valued Borel measure (POVM) $M(dx)$ on $ \mathcal{X}$ such
that
\begin{equation}
\Phi (\rho )=\int\limits_{\mathcal{X}}\rho ^{\prime }(x)\mu _{\rho }(dx),
\label{thm}
\end{equation}
where $\mu _{\rho }(B)=\mathrm{Tr}\rho M(B)$ for all Borel $B\subseteq
\mathcal{X}$.}

\textbf{Proof.} Notice first that conditions 1),2) in the definition of
channel are readily verified for the map (\ref{thm}). Let us show that the
channel (\ref{thm}) is entanglement-breaking. Let $\omega \in \mathfrak{S}(
\mathcal{H}\otimes \mathcal{K})$, where $\mathcal{K}$ is a finite
dimensional Hilbert space. We have
\begin{equation}
(\Phi \otimes \mathrm{Id}_{\mathcal{K}})(\omega )=\int\limits_{\mathcal{X}
}\rho ^{\prime }(x)\otimes m_{\omega }(dx),  \label{ep}
\end{equation}
where
\begin{equation*}
m_{\omega }(B)=\mathrm{Tr}_{\mathcal{H}}\omega (M(B)\otimes I_{\mathcal{K}
}),\quad B\subseteq \mathcal{X}.
\end{equation*}
It is easy to see that any matrix element of $m_{\omega }$ (in a particular
basis) is a complex valued measure on $\mathcal{X}$ absolutely continuous
with respect to the probability measure $\mu _{\omega }(B)=\mathrm{Tr}
m_{\omega }(B),\;B\subseteq \mathcal{X}$. The Radon-Nikodym theorem implies
representation
\begin{equation*}
m_{\omega }(B)=\int\limits_{B}\sigma _{\omega }(x)\mu _{\omega }(dx),
\end{equation*}
where $\sigma _{\omega }(x)$ is a function on $\mathcal{X}$ taking
values in $\mathfrak{S}(\mathcal{K}).$ By using this
representation we can rewrite (\ref{ep}) as
\begin{equation}  \label{epp}
(\Phi \otimes \mathrm{Id}_{\mathcal{K}})(\omega
)=\int\limits_{\mathcal{X} }\rho ^{\prime }(x)\otimes \sigma
_{\omega }(x)\mu _{\omega }(dx),
\end{equation}
which reduces to (\ref{sepa}) by change of variables and hence is
separable by Lemma 1.

Conversely, let $\Phi $ be an entanglement-breaking channel. Fix a
state $ \sigma $ in $\mathfrak{S}(\mathcal{H})$ of full rank and
let $\{|i\rangle \}_{i=1}^{+\infty }$ be the basis of eigenvectors
of $\sigma $ with the corresponding (positive) eigenvalues
$\{\lambda _{i}\}_{i=1}^{+\infty }$. Consider the vector
\begin{equation*}
|\Omega \rangle =\sum_{i=1}^{+\infty }\lambda _{i}^{1/2}|i\rangle \otimes
|i\rangle
\end{equation*}
in the space $\mathcal{H}\otimes \mathcal{H}$. Since $\Phi $ is
entanglement-breaking, the state
\begin{equation}
\rho =(\mathrm{Id}_{\mathcal{H}}\otimes \Phi )(|\Omega \rangle \langle
\Omega |)  \label{out}
\end{equation}
in $\mathfrak{S}(\mathcal{H}\otimes \mathcal{H}^{\prime })$ is
separable. By (\ref{sep}) there exists a probability measure $\nu $
on $\mathfrak{P}( \mathcal{H})\times
\mathfrak{P}(\mathcal{H}^{\prime })$ such that
\begin{equation}
(\mathrm{Id}_{\mathcal{H}}\otimes \Phi )(|\Omega \rangle \langle
\Omega
|)=\int\limits_{\mathfrak{P}(\mathcal{H})}\int\limits_{\mathfrak{P}(\mathcal{
H}^{\prime })}|\varphi \rangle \langle \varphi |\otimes |\psi
\rangle \langle \psi |\nu (d\varphi d\psi ).  \label{one}
\end{equation}
This implies
\begin{eqnarray}
\sigma  &=&\mathrm{Tr}_{\mathcal{H}^{\prime }}(Id_{\mathcal{H}}\otimes \Phi
)(|\Omega \rangle \langle \Omega |)  \notag \\
&=&\int\limits_{\mathfrak{P}(\mathcal{H})}\int\limits_{\mathfrak{P}(\mathcal{
H}^{\prime })}|\varphi \rangle \langle \varphi |\nu (d\varphi d\psi
)  \notag
\\
&=&\int\limits_{\mathfrak{P}(\mathcal{H})}\int\limits_{\mathfrak{P}(\mathcal{
H}^{\prime })}|\bar{\varphi}\rangle \langle \bar{\varphi}|\nu
(d\varphi d\psi ),  \label{ppp}
\end{eqnarray}
where the bar denotes complex conjugation in the basis$\{|i\rangle
\}_{i=1}^{+\infty }.$ By this equality for arbitrary Borel $B\subseteq
\mathfrak{P}(\mathcal{H}^{\prime })$ the operator
\begin{equation*}
M(B)=\sigma ^{-1/2}\left[ \int\limits_{\mathfrak{P}(\mathcal{H}
)}\,\int\limits_{B}|\bar{\varphi}\rangle \langle \bar{\varphi}|\nu
(d\varphi d\psi )\,\right] \sigma ^{-1/2}
\end{equation*}
can be defined as a bounded positive operator on $\mathcal{H}$ such
that $ M(B)\leq M(\mathcal{X})=I_{\mathcal{H}}$. It is easy to see
that $M(d\psi )$ is a POVM on
$\mathcal{X}=\mathfrak{P}(\mathcal{H}^{\prime })$.

Consider the entanglement-breaking channel
\begin{equation*}
\hat{\Phi}(\rho )=\int\limits_{\mathfrak{P}(\mathcal{H}^{\prime })}|\psi
\rangle \langle \psi |\mu _{\rho }(d\psi ),
\end{equation*}
where $\mu _{\rho }$ is the Borel probability measure defined by $\mu _{\rho
}(B)=\mathrm{Tr}\rho M(B),B\subseteq \mathcal{X}$. To prove that $\Phi (\rho
)=\hat{\Phi}(\rho ),$ it is sufficient to show that
\begin{equation*}
\hat{\Phi}(|i\rangle \langle j|)=\Phi (|i\rangle \langle j|)
\end{equation*}
for all $i,j.$ But
\begin{equation*}
\begin{array}{c}
\hat{\Phi}(|i\rangle \langle
j|)=\int\limits_{\mathfrak{P}(\mathcal{H} ^{\prime })}|\psi
\rangle \langle \psi |\langle j|M(d\psi )|i\rangle  \\ =\lambda
_{i}^{-1/2}\lambda
_{j}^{-1/2}\int\limits_{\mathfrak{P}(\mathcal{H}
)}\int\limits_{\mathfrak{P}(\mathcal{H}^{\prime })}\langle
i|\varphi \rangle \langle \varphi |j\rangle |\psi \rangle \langle
\psi |\nu (d\varphi d\psi )=\Phi (e_{ij}),
\end{array}
\end{equation*}
where
\begin{equation*}
e_{ij}=\lambda _{i}^{-1/2}\lambda
_{j}^{-1/2}\mathrm{Tr}_{\mathcal{H} }(|j\rangle \langle i|\otimes
I)|\Omega \rangle \langle \Omega |=|i\rangle \langle j|.\quad
\square
\end{equation*}

The representation (\ref{thm}) is by no means unique. A natural
question is: for an arbitrary entanglement-breaking channel, is it
possible to find a representation (\ref{thm}) with a purely atomic
POVM?\ From the proof of the theorem one can see that this is the
case if and only if the state (\ref{out}) is countably decomposable,
hence, as we show in the next section, the answer is negative. This
implies existence of an entanglement-breaking channel, which has no
Kraus representation with operators of rank 1, in contrast to the
finite dimensional case \cite{R}. Indeed, it is easy to see that
such a Kraus representation with operators of rank 1 is equivalent
to a representation (\ref{thm}) of this channel with a purely atomic
POVM.

\section{Example}

We shall consider the one-dimensional rotation group represented as the
interval $[0,2\pi )$ with addition \textrm{mod}$2\pi .$ Let $\mathcal{H=}
L^{2}$ $[0,2\pi )$ with the normalized Lebesgue measure $\frac{dx}{2\pi },$
and let $\left\{ |k\rangle ;k\in \mathbf{Z}\right\} $ be the orthonormal
basis of trigonometric functions, so that
\begin{equation*}
\langle k|\psi \rangle =\int_{0}^{2\pi }e^{-ixk}\psi (x)\frac{dx}{2\pi }.
\end{equation*}
Consider the unitary representation $x\rightarrow V_{x}$ , where $
V_{x}=\sum_{-\infty }^{\infty }e^{ixk}|k\rangle \langle k|,$ so
that $ (V_{u}\psi )(x)=\psi (x-u)$. For any fixed state vector
$|\varphi \rangle \in \mathcal{H},$ the formula
\begin{equation}
\Phi (\rho )=\int_{0}^{2\pi }V_{x}|\varphi \rangle \langle \varphi
|V_{x}^{\ast }\mu _{\rho }(dx),  \label{channel}
\end{equation}
where $\mu _{\rho }(B)=\mathrm{Tr}\rho E(B),$ and $E(dx)$ is the spectral
measure of the operator of multiplication by $x$ in $\mathcal{H=}
L^{2}[0,2\pi )$, defines entanglement--breaking channel. The channel $\Phi $
is rotation-covariant in that
\begin{equation}  \label{cov-prop}
\Phi (V_{x}\rho V_{x}^{\ast })=V_{x}\Phi (\rho )V_{x}^{\ast },\quad \ x\in
\lbrack 0,2\pi ).
\end{equation}
It is not difficult to check that
\begin{equation}  \label{mu-exp}
\mu _{\rho }(B)=\int_{B}\langle x|\rho |x\rangle \frac{dx}{2\pi },
\end{equation}
where $\langle x|\rho |x\rangle =p(x)$ is the diagonal value of the
density operator $\rho $ which is unambiguously defined as a
probability density in $ L^{1}$.

\textbf{Theorem 3.} \textit{For arbitrary state vectors }$|\varphi
_{j}\rangle \in \mathcal{H}_{j}\simeq L^{2}$\textit{$[0;2\pi );j=1,2,$ with
nonvanishing Fourier coefficients the separable state
\begin{equation}
\rho _{12}=\int_{0}^{2\pi }V_{x}^{(1)}|\varphi _{1}\rangle \langle \varphi
_{1}|V_{x}^{(1)\ast }\otimes V_{x}^{(2)}|\varphi _{2}\rangle \langle \varphi
_{2}|V_{x}^{(2)\ast }\;\frac{dx}{2\pi }  \label{cind}
\end{equation}
in }$\mathcal{H}_{1}\otimes \mathcal{H}_{2}$ \textit{is not countably
decomposable.}

\textbf{Proof.} Suppose $\rho $ is countably decomposable, then by
(\ref{nec}) there exist nonzero \textit{$\alpha _{j}\in
\mathcal{H}_{j}$} such that
\begin{equation}
\rho _{12}\geq |\alpha _{1}\rangle \langle \alpha _{1}|\otimes |\alpha
_{2}\rangle \langle \alpha _{2}|.  \label{main-ineq}
\end{equation}
Taking partial traces we obtain
\begin{equation*}
\int_{0}^{2\pi }V_{x}^{(j)}|\varphi _{j}\rangle \langle \varphi
_{j}|V_{x}^{(j)\ast }\;\frac{dx}{2\pi }\geq |\alpha _{j}\rangle \langle
\alpha _{j}|,
\end{equation*}
whence
\begin{equation}
|\langle k|\varphi _{j}\rangle |\geq |\langle k|\alpha _{j}\rangle |;\quad
j=1,2;\quad k\in \mathbb{Z}.  \label{F-c-ineq}
\end{equation}

Inequality (\ref{main-ineq}) means that
\begin{equation}
\int_{0}^{2\pi }\left\vert \langle \lambda _{1}|V_{x}^{(1)}|\varphi
_{1}\rangle \right\vert ^{2}\left\vert \langle \lambda
_{2}|V_{x}^{(2)}|\varphi _{2}\rangle \right\vert ^{2}\;\frac{dx}{2\pi }\geq
|\langle \lambda _{1}|\alpha _{1}\rangle |^{2}|\langle \lambda _{2}|\alpha
_{2}\rangle |^{2}  \label{b-ineq}
\end{equation}
for arbitrary $\lambda _{j}\in L^{2}[0;2\pi )$. For technical convenience we
will assume that the functions $\lambda _{j}(x)$ have finite number of
nonzero Fourier coefficients. Introducing
\begin{equation*}
\mu _{j}(x)=\langle \lambda _{j}|V_{x}^{(j)}|\varphi _{j}\rangle
=\sum_{k=-\infty }^{+\infty }\langle \lambda _{j}|k\rangle \langle k|\varphi
_{j}\rangle e^{ikx},
\end{equation*}
so that $\langle k|\mu _{j}\rangle =\langle \lambda _{j}|k\rangle \langle
k|\varphi _{j}\rangle ,$ we see that $\mu _{j}\mapsto \langle \lambda
_{j}|\alpha _{j}\rangle $ are linear functionals of $\mu _{j}$ running over
the subspace of trigonometric polynomials in $L^{2}[0;2\pi )$. These
functional are in fact continuous. Indeed, choosing $k$ such that $\langle
k|\alpha _{2}\rangle \neq 0,$ we find from (\ref{b-ineq})
\begin{equation*}
|\langle \lambda _{1}|\alpha _{1}\rangle |^{2}\leq \left\vert \frac{\langle
k|\varphi _{2}\rangle }{\langle k|\alpha _{2}\rangle }\right\vert
^{2}\int_{0}^{2\pi }\left\vert \mu _{1}(x)\right\vert ^{2}\;\frac{dx}{2\pi }
\end{equation*}
for all trigonometric polynomials $\mu _{1}.$ Hence by Riesz theorem there
exists $\beta _{1}\in L^{2}[0;2\pi )$ such that $\langle \lambda _{1}|\alpha
_{1}\rangle =\langle \beta _{1}|\mu _{1}\rangle .$ Applying similar
reasoning to $j=2,$ we can transform (\ref{b-ineq}) to the form
\begin{equation}
|\langle \beta _{1}|\mu _{1}\rangle |^{2}|\langle \beta _{2}|\mu _{2}\rangle
|^{2}\leq \int_{0}^{2\pi }\left\vert \mu _{1}(x)\mu _{2}(x)\right\vert
^{2}\; \frac{dx}{2\pi },  \label{b12}
\end{equation}
where $\beta _{j}\in L^{2}[0;2\pi ).$

Now we can extend the inequality (\ref{b12}) to more general
functions $ \mu_j $ for which both sides of this inequality are
defined e. g. measurable a. e. uniformly bounded functions on
$[0;2\pi ]$. The characteristic functions of intervals belong to
this class, and so is a dense set of functions with support in any
specified interval.

Consider a partitioning of the interval $[0,2\pi]$ into intervals of
length $ \leq\varepsilon$, and pick one of these intervals, say
$I_2$, on which $ \beta_2$ is not a.e. zero. Then we can find an
admissible function $\mu_2$ supported in $I_2$ such that $\langle
\beta _{2}|\mu _{2}\rangle\neq 0$. But then, for any $\mu_1$
supported on the complement of $I_2$, the right hand side of
(\ref{b12}) vanishes, and therefore $\beta_1$ vanishes a.e. on the
complement of $I_2$. It follows that the support of $\beta_1$ has
measure $ \leq\varepsilon$ for all $\varepsilon$, i.e., $\beta_1$
vanishes a.e., and hence $\alpha_1=0$. $\square $

Let $\mathfrak{W}$ be the subset of all separable states which are not
countably decomposable.

\textbf{Corollary.} \textit{An arbitrary pure product state in
$\mathfrak{S} ( \mathcal{H}\otimes\mathcal{K}$) can be approximated
by a sequence from $ \mathfrak{W}$.}

\textbf{Proof.} We can take $\mathcal{H},\mathcal{K\simeq }L^{2}[0;2\pi ).$
Since an arbitrary function in $L^{2}$ can be approximated by functions with
nonzero Fourier coefficients it sufficient to consider pure product state $
|\varphi _{1}\rangle \langle \varphi _{1}|\otimes |\varphi _{2}\rangle
\langle \varphi _{2}|$, where $\varphi _{j}(x)$ have this property.

Consider the sequence of states
\begin{equation*}
\rho _{12}^{(n)}=\int_{0}^{2\pi /n}V_{x}^{(1)}|\varphi _{1}\rangle \langle
\varphi _{1}|V_{x}^{(1)\ast }\otimes V_{x}^{(2)}|\varphi _{2}\rangle \langle
\varphi _{2}|V_{x}^{(2)\ast }\;\frac{ndx}{2\pi },
\end{equation*}
such that $\rho _{12}^{(n)}\rightarrow |\varphi _{1}\rangle \langle \varphi
_{1}|\otimes |\varphi _{2}\rangle \langle \varphi _{2}|$ as $n\rightarrow
\infty .$ For the state $\rho _{12}$ given by the (\ref{cind}), we have
\begin{equation*}
\rho _{12}=\frac{1}{n}\sum_{k=0}^{n-1}\left(V_{\frac{2\pi
k}{n}}^{(1)}\otimes V_{ \frac{2\pi k}{n}}^{(2)}\right)\rho
_{12}^{(n)}\left(V_{\frac{2\pi k}{n}}^{(1)}\otimes V_{\frac{2\pi
k}{n}}^{(2)}\right)^{\ast},
\end{equation*}
therefore $\rho _{12}^{(n)}\in \mathfrak{W}$. Indeed, otherwise we could
construct a countable decomposition for $\rho _{12}$. $\square $

\textbf{Conjecture.} \textit{The subset $\mathfrak{W}$ is dense in the set
of all separable states.}

\section{The classical capacity}

In terms of the relative entropy, the $\chi -$capacity of an arbitrary
quantum channel $\Phi $ is defined by the relation
\begin{equation}
C(\Phi )=\sup_{\{\pi _{i},\rho _{i}\}}\sum_{i}\pi _{i}H(\Phi (\rho
_{i});\Phi (\bar{\rho})),  \label{chicap}
\end{equation}
where supremum is over all (finite) ensembles $\{\pi _{i},\rho _{i}\}$ with
the average $\bar{\rho}$. For entanglement-breaking channels the $\chi$-
capacity is additive (for the infinite-dimensional case see \cite{Sh-2}),
hence it gives the classical capacity of the channel.

\textbf{Theorem 4. } \textit{The classical capacity of the channel
(\ref {channel}) is equal to}
\begin{equation}
C(\Phi)=-\sum_{k=-\infty }^{\infty }|\langle k|\varphi \rangle |^{2}\log
|\langle k|\varphi \rangle |^{2}.  \label{cap}
\end{equation}

\textbf{Proof.} Let us first show that the closure of
$\mathrm{ran}\Phi =\Phi \left(\mathfrak{S}(\mathcal{H})\right) $
coincides with the set
\begin{equation}  \label{set}
\overline{\mathrm{co}}\left\{ V_{x}|\varphi \rangle \langle \varphi
|V_{x}^{\ast };x\in \lbrack 0,2\pi ]\right\}.
\end{equation}
By (\ref{mu-exp}) and using the fact that arbitrary probability
density $ p(\cdot )$ can be obtained as the diagonal value of a
density operator (in fact, of a pure state), we have
\begin{equation}\label{ran}
\mathrm{ran}\Phi =\left\{ \int_{0}^{2\pi }V_{x}|\varphi \rangle \langle
\varphi |V_{x}^{\ast }p(x)\frac{dx}{2\pi };\quad p(\cdot )\in \mathcal{P}
\right\} ,
\end{equation}
where $\mathcal{P}$ is the convex set of probability densities on
$[0,2\pi ]$. The set of absolutely continuous probability measures
is weakly dense in the Choquet simplex $\mathcal{M}$ of all Borel
probability measures on $[0,2\pi ]$, while the map
$$
\mu\mapsto\int_{0}^{2\pi }V_{x}|\varphi \rangle \langle \varphi
|V_{x}^{\ast }\mu (dx)
$$
is continuous (due to the aforementioned result in \cite{Dev},
\cite{D-A}), therefore the set (\ref{ran}) is dense in the compact
convex set
\begin{equation*}
\left\{ \int_{0}^{2\pi }V_{x}|\varphi \rangle \langle \varphi
|V_{x}^{\ast }\mu (dx);\;\mu \in \mathcal{M}\right\},
\end{equation*}
which coincides with the set (\ref{set}) by lemma 1.

Now suppose $C(\Phi )$ is finite. By proposition 1 in \cite{Sh-2} there
exists the unique state $\Omega (\Phi )$ in $\overline{\mathrm{ran}}\Phi $
such that
\begin{equation}
C(\Phi )=\sup_{\rho \in \mathfrak{S}(\mathcal{H})}H(\Phi (\rho );\Omega
(\Phi )).  \label{sup}
\end{equation}
The uniqueness of this state and the covariance (\ref{cov-prop})
imply $V_{x}\Omega (\Phi )V_{x}^{\ast }=\Omega (\Phi )$ for all $x$
and hence
\begin{equation*}
\Omega (\Phi )=\int_{0}^{2\pi }V_{x}|\varphi \rangle \langle \varphi
|V_{x}^{\ast }\frac{dx}{2\pi }=-\sum_{k=-\infty }^{\infty }|\langle
k|\varphi \rangle |^{2}|k\rangle \langle k|.
\end{equation*}
Since $\mathrm{ran}\Phi $ is dense in the set (\ref{set}), and the
relative entropy is lower semicontinuous and convex, we see that the
supremum (\ref {sup}) is equal to $H(|\varphi \rangle \langle
\varphi |;\Omega (\Phi ))$, so we have
\begin{equation*}
C(\Phi )=-\langle \varphi |\log \Omega (\Phi )|\varphi \rangle
=H\left( \Omega (\Phi )\right)
\end{equation*}
which is equal to the right side of (\ref{cap}).

To complete the proof it is sufficient to show that
\begin{equation}
-\sum_{k=-\infty }^{\infty }|\langle k|\varphi \rangle |^{2}\log |\langle
k|\varphi \rangle |^{2}<\infty  \label{series}
\end{equation}
implies finiteness of the capacity $C(\Phi )$. Let us show first
that (\ref {series}) implies continuity of the output entropy
$H(\Phi (\rho )).$ Indeed, assuming (\ref{series}) one can find a
sequence $\left\{ h_{k}\right\} $ of positive numbers such that
$h_{k}\uparrow +\infty $ with $ |k|\uparrow +\infty $ and
\begin{equation*}
\sum_{k=-\infty }^{+\infty }h_{k}|\langle k|\varphi \rangle |^{2}(-\log
|\langle k|\varphi \rangle |^{2})\equiv h<\infty .
\end{equation*}
Introducing selfadjoint operator
\begin{equation*}
H=\sum_{k=-\infty }^{+\infty }h_{k}(-\log |\langle k|\varphi \rangle
|^{2})|k\rangle \langle k|,
\end{equation*}
we have
\begin{equation}
\mathrm{Tr}\exp (-\beta H)=\sum_{k=-\infty }^{+\infty }|\langle k|\varphi
\rangle |^{2\beta h_{k}}<\infty ;\quad \beta >0,  \label{e1}
\end{equation}
since $\beta h_{k}\geq 1$ for all sufficiently large $k.$ By (\ref{ran}),
for arbitrary $\rho \in \overline{\mathrm{ran}}\Phi $ there is a probability
measure $\mu $ such that $\rho =\int_{0}^{2\pi }V_{x}|\varphi \rangle
\langle \varphi |V_{x}^{\ast }\mu (dx)$. Thus
\begin{equation}
\mathrm{Tr}\rho H=\sum_{k=-\infty }^{+\infty }h_{k}|\langle k|\varphi
\rangle |^{2}(-\log |\langle k|\varphi \rangle |^{2})\int_{0}^{2\pi }\mu
(dx)=h  \label{e2}
\end{equation}
for $\rho \in \overline{\mathrm{ran}}\Phi$. It is well known
\cite{wehrl} that the relations (\ref{e1}) and (\ref{e2}) imply
continuity of the restriction of the quantum entropy to the set
$\overline{\mathrm{ran}}\Phi $ , which implies continuity of the
output entropy $H(\Phi (\rho ))$ on $ \mathfrak{S}(\mathcal{H})$.

Now the maximum of the quantum entropy on the set $\overline{\mathrm{ran}}
\Phi $ is attained on the state $\Omega (\Phi )$ and is equal to the sum of
the series (\ref{series}). Indeed, by the unitary invariance, concavity and
continuity of the output entropy, we have
\begin{equation*}
H\left( \rho \prime \right) =\int_{0}^{2\pi }H\left( V_{x}\rho \prime
V_{x}^{\ast }\right) \frac{dx}{2\pi }\leq H\left( \int_{0}^{2\pi }V_{x}\rho
\prime V_{x}^{\ast }\frac{dx}{2\pi }\right) =H\left( \Omega (\Phi )\right)
\end{equation*}
for any $\rho \prime \in \overline{\mathrm{ran}}\Phi $. Therefore
\begin{equation*}
C(\Phi )\leq \max_{\rho }H\left( \Phi (\rho )\right) =H\left( \Omega (\Phi
)\right) <\infty.
\end{equation*}
Moreover, continuity of the entropy on the set $\overline{
\mathrm{ran}}\Phi$, which is the convex closure of a family of pure
states, implies $\inf_{\rho}H(\Phi(\rho))=0 $ and $C(\Phi
)=\max_{\rho }H\left( \Phi (\rho )\right). \quad\square$

\end{document}